# The Interface between Titanium and Carbon Nanotubes

Alexandre Felten*[a], Irene Suarez-Martinez*[b], Xiaoxing Ke[c], Gustaaf Van Tendeloo[c], Jacques Ghijsen[a], Jean-Jacques Pireaux[a], Wolgang Drube[d], Carla Bittencourt[e] and Christopher P. Ewels[b]

*((Dedication, optional))*

We study the interface between carbon nanotubes (CNTs) and surface deposited titanium using electron microscopy and photoemission spectroscopy, supported by density functional calculations. Charge transfer from the Ti atoms to the nanotube and carbide formation is observed at the interface which indicates strong interaction. Nevertheless, the presence of oxygen between the Ti and the CNTs significantly weakens the Ti-CNT interaction. Ti atoms at the surface will preferentially bond to oxygenated sites. Potential sources of oxygen impurities are examined, namely oxygen from any residual atmosphere and pre-existing oxygen impurities on the nanotube surface, which we enhance through oxygen plasma surface pre-treatment. Variation in literature data concerning Ohmic contacts between Ti and carbon nanotubes is explained via sample pre-treatment and differing vacuum levels, and we suggest improved treatment routes for reliable Schottky barrier free Ti-nanotube contact formation.

## Introduction

Carbon nanotubes appear as a good candidate to replace state-of-the-art inlaid copper interconnects in microprocessors. A key step for integration of CNTs in actual devices is the formation of stable and low-resistance ohmic contacts. Experimental results and calculations seem to agree that Ti and Pd are the best candidates for contacts in carbon nanotube based devices[1-8]. They are superior to conventional metals previously used in electronics such as Au, Al, and Pt which have associated high resistance Schottky barriers. Theoretical studies of pure metals on graphite suggest Ti as the best metal for contacts followed by Pd[5], yet while low resistance Ohmic contacts have been experimentally reported in the case of Ti[1-3], experimental results give Pd as the best choice due to better reliability and reproducibility[2,6]. One of the reasons invoked for this discrepancy is the high chemical reactivity of Ti compared to the other metals [2,6,9]: in realistic systems titanium oxidation occurs. But until now, no chemical study of the Ti/nanotube interface has been carried out in order to understand the phenomenon, nor supporting theoretical calculations.

In this paper, we investigate the growth, electronic structure and chemical composition of Ti atoms on multiwall carbon nanotubes (MWNTs) by transmission electron microscopy (TEM), core levels photoemission spectroscopy (XPS) and density functional calculations.

## Results and Discussion

Evaporation of Ti on pristine MWCNTs leads to continuous amorphous coverage of the nanotube surface, even for small amounts of evaporated metal. High resolution TEM images in Figure 1 show that the film is continuous from at least 1 nm coverage; unlike for example Pd which shows isolated particle formation.

Calculations for small Ti clusters confirm the tendency of Ti to wet the carbon surface. A single Ti atom strongly binds to graphene (2.64 eV), with Ti above a hexagon centre with short Ti-C bonds (2.24 Å), in good agreement with previous studies[10]. We have examined a range of 2D and 3D structures for $Ti_n$ on graphene (n=1-4). For each cluster size the most stable structure is planar parallel to the graphene (see Figure 2). This is in contrast to Ti$_4$ in the gas phase where we find the most stable form is tetrahedral, in agreement with previous Ti cluster calculations[11]. A continuous monolayer of epitaxial Ti forms a buckled layer on average 2.0 Å above the graphitic layer, with Ti-C distances of 2.13 Å or 2.39 Å depending on whether the Ti atom lies above C atoms or hexagon centres respectively. In this case our binding energy is 5.43 eV/Ti atom. Our calculated barrier for Ti diffusion on the graphitic surface is 0.75 eV, as compared to ~0.1eV for Au and Pd. Thus while a simple first order Arhennius type equation with the Debye attempt frequency ($10^{13}$ Hz) suggests that isolated Au or Pd atoms on the graphitic surface at room temperature will be moving at more than $2\times10^{11}$ hops/second, Ti atoms will only be moving at 2 hops/second, i.e. Ti atoms are almost immobile. This is consistent with the Ti coating being amorphous whereas Pd and Au particles are crystalline after deposition[12,13].

[a]  Dr A. Felten, Dr J. Ghijsen, Prof. Dr. J.-J. Pireaux
Centre de Recherche en Physique de la Matière et du Rayonnement (PMR)
University of Namur
61 rue de Bruxelles, 5000, Namur, Belgium
Fax: +32 81 724595
E-mail: alexandre.felten@fundp.ac.be

[b]  Dr I. Suarez-Martinez, Dr. C.P. Ewels
Institut des Matériaux de Nantes (IMN)
CNRS UMR6502
2 rue de la Houssiniere, BP 32229, Nantes, France
E-mail: irene.suarez@cnrs-imn.fr

[c]  X. Ke, Prof. Dr. G. Van Tendeloo
EMAT
University of Antwerp
Groenenborgerlaan 171, B2020 Antwerpen, Belgium

[d]  Dr. W. Drube
Deutschen Elektronen-Synchrotron DESY
Notkestrasse 85, D-22603 Hamburg, Germany

[e]  Dr. C. Bittencourt
LCIA
University of Mons-Hainaut
Parc Initialis, Av. Nicolas Copernic 1, Mons 7000, Belgium

Introducing O$_2$ strongly modifies the interaction of Ti with the graphitic layer. The two O atoms bind strongly to the single Ti atom (8.59 eV/O$_2$, Ti-O bonds of 1.66 Å), which significantly weakens the Ti-C interaction (bonds dilated to 2.60 Å). Essentially the TiO$_2$ unit becomes molecular and sits above the graphene π-cloud, again in agreement with previous calculations[10]. The presence of oxygen significantly weakens the Ti-C interaction.

The calculations also show strong qualitative variations in charge transfer behaviour. An isolated Ti atom on graphene has a net spin of 2.41$\mu_B$, and Mulliken distribution analysis shows a strong net charge transfer from the Ti to the graphene layer resulting in a Ti charge of +0.98e. On addition of oxygen the Ti charge increases to +1.36e, however due to the charge on the oxygen atoms (-0.83e/O) the result is a much weaker net charge transfer *in the opposite direction* from the graphene to the TiO$_2$ unit of 0.29e.

Thus we clearly see that the presence of oxygen *at the interface* between Ti and nanotube critically controls the extent of the Ti-C interaction. Blackstock et al.[14] recently investigated a similar problem at Ti/organic monolayer interfaces. They observed a formation of 1-2 nm thick TiO$_2$ layer at the interface and attributed it to the vacuum conditions, surface water and oxygen diffusion from the metal oxide substrate. We investigate experimentally the importance of oxygen contamination from two potential sources: residual atmosphere still present in the chamber during Ti evaporation or from oxidized defects present on the nanotube surface.

We first investigate the deposition at ultra-high vacuum of $10^{-9}$ mbar. XPS of pristine nanotubes shows trace oxygen present at less than 1 at%. After the first step of titanium evaporation (sample 1), the Ti $2p_{3/2}$ peak has a binding energy of ~459.0 eV (see Figure 3b), which is characteristic of Ti$^{4+}$ present in TiO$_2$[15,16]. This suggests that titanium atoms first react with any available oxygen to form completely oxidized titanium. As the amount of deposited metal increases, the Ti $2p_{3/2}$ peak evolves into a broad structure due to the presence of Ti in different chemical environments. First, it shows structures ranging from 455 to 460 eV corresponding to oxides with different stoichiometries: peaks at 459.0 eV, 457.6 eV, 456.7 eV and 455.3 eV are reported as TiO$_2$, Ti$_3$O$_5$, Ti$_2$O$_3$ and TiO respectively[15]. Notably with increasing Ti thickness there is a shift in the TiO$_x$ peaks towards increasingly oxygen depleted phases, consistent with the presence of a fixed reservoir of oxygen shared by an increasing quantity of Ti. A new peak attributed to Ti atoms bound to carbon progressively appears at lower binding energy (454.8 eV). From sample 5 (10 Å of Ti evaporated), this peak becomes predominant. Finally, for a thick layer of Ti (sample 7), a peak at 453.8 eV is observed corresponding to metallic Ti. This evolution with the thickness is consistent with the theoretical modelling showing a clear preference for Ti to form Ti-O bonds, followed by Ti-C and finally Ti-Ti.

The evolution of the C $1s$ spectra with increasing Ti coverage is shown in Figure 3a. At increasing Ti coverage, simultaneously with the appearance of a peak at 281.9 eV associated with the formation of Ti-C bonds, a shoulder on the main C $1s$ peak appears near 285.5 eV. This shoulder is due to charge transfer from the increasing Ti layer to the nanotube. We can analyze this effect further by subtracting from the spectra the contribution of photoelectrons emitted from the inner walls, assuming the layers are electronically isolated, *i.e.* all charge transfer from the Ti occurs to the outer walls of the CNT[17]. The subtracted spectra are shown in Figure 4. Assuming a rigid band shift, peak displacement to higher energies corresponds to increased charge transfer from Ti to the carbon. The peaks form two distinct families, those at low Ti coverage (samples 1, 2) which are only weakly asymmetric, and those at higher Ti coverage (samples 3-7) which are highly asymmetric with a strong high energy shoulder. This asymmetry has been associated with electron-hole pair excitations[18,19]: the potential created between the photohole and the remaining electrons after the photoemission induces the promotion of electrons near the Fermi level to empty states just above it. The increased asymmetry is thus consistent with an increase in the DOS associated with charge transfer. The transition from weak to strong charge transfer between samples 2 and 3 corresponds with the first appearance of Ti-C interaction in the Ti $2p$ spectrum of sample 3 and the end of oxide formation and is in agreement with the calculations showing higher charge transfer for Ti bonded directly to carbon than with oxygen present. We note that for thick Ti films we no longer see any increase in charge transfer in the subtracted C $1s$ spectra, *i.e.* the top Ti metal does not contribute charge to the underlying nanotube.

In order to investigate the importance of the residual atmosphere, we next deposited a 10 Å layer of Ti under three different pressures: $10^{-6}$, $10^{-7}$ and $10^{-8}$ mbar, using Au films as the substrate. Figure 5 shows the titanium oxide content in the deposited layers as calculated from a deconvolution of the Ti $2p_{3/2}$ XPS peak area into metal, carbide and oxide components. Oxygen is always present in the titanium film. At $10^{-6}$ mbar, the layer is fully oxidized with a Ti $2p$ peak showing a large majority of TiO$_2$. Improving the vacuum leads to a decrease in the oxide content with a drastic decrease of the TiO$_2$ peak. At $10^{-8}$ mbar, half of the Ti atoms are oxidized with only 10% forming TiO$_2$, the rest forming TiO$_x$ (x<2) phases.

Another potential source of oxygen contamination is oxygen containing defects on the nanotube surfaces. In order to investigate the importance of these, we oxygen plasma treated the MWNTs before the Ti is deposited. XPS showed that 15 at% of oxygen is grafted to the nanotube surface using the oxygen plasma treatment. Figure 6a shows the comparison between the C 1s peaks recorded from pristine and oxygen plasma-treated nanotubes. Photoelectrons emitted from carbon atoms in the "graphite-like" walls generate the main feature of these spectra at binding energy of 284.3 eV. The chemical modification produced by the plasma treatment is identified by a broad structure peaking at 287.5 eV: it is attributed[20] to photoelectrons emitted from carbon atoms belonging to oxygen groups (hydroxyl, carbonyl and/or carboxyl). The evaporation of titanium was then performed simultaneously on pristine and oxygen plasma treated MWCNTs. Compared to the spectra recorded on pristine nanotubes, similar trends are observed but with some notable differences: (1) sample 3, which showed components due to both Ti-C and Ti-O bonds in the pristine case, now only shows a component due to Ti-O bonds (Figure 7), which is consistent with a larger available reservoir of oxygen (2) the C *1s* level of functionalized nanotubes shows a peak around 286.7 eV, corresponding to oxygen grafted to carbon during the oxygen plasma treatment. This peak disappears as the amount of Ti increases (Figure 6b); (3) the asymmetry of the C *1s* peak is less pronounced compared to non-treated nanotubes. This indicates that due to Ti oxide formation, less charge is transferred to the oxygen treated nanotube than to the pristine one, in agreement with the calculations. For example, at Ti thickness of 2 Å, the FWHM of the C 1s peak after background subtraction is 1.65 eV for pristine MWCNTs and 1.15 eV for the functionalized ones (Figure 6*c)*.

Oxygen plasma treatment under similar conditions induces a variety of defects into the nanotube wall as identifies by XPS[20], primarily aldehyde/ketones (>C=O) but also ester/carboxyls (-COO-)[21] for longer treatment times. Since there are many possible defect configurations giving rise to these systems, we modelled the behaviour of single Ti atoms in the vicinity of an oxygenated vacancy (VacO$_2$) as the smallest test system, which contains oxygen atoms in both ketone and ether bonded configurations. Several metastable configurations for the addition of a single Ti atom are possible with similar energies, and in all cases the Ti binds strongly (>5 eV), i.e. >3 eV more strongly than to the pristine carbon sheet. Some of the lowest energy configurations involve oxygen atoms scavanged from the vacancy by the Ti atom, however we anticipate the barrier to break the C-O bond will be significant and such configurations will be unobtainable experimentally at room temperature. All other stable configurations involve Ti in the immediate viscinity of the VacO$_2$ structure, with bonds between the Ti atom and both oxygen and carbon atoms. The high binding energies show that once trapped by an oxygenated surface defect Ti will clearly be immobile at room temperature. The calculations therefore confirm that oxygen plasma induced surface defects will be strong trapping sites for Ti, in agreement with experiment.

## Conclusion

The quantity of oxygen present in the *initial surface layer* crucially controls the Ti-nanotube interaction. Later surface oxygenation of the Ti particle is less important for the contact behaviour. This is because initial Ti deposition preferentially forms Ti-O bonds, strongly reducing the interaction between the Ti and the nanotube. Once there is no oxygen available the remaining Ti forms Ti-C bonds with strong interaction as seen in the large charge transfer.

The quality of the Ti-CNT electrical contact can be determined by the quality of vacuum under which it was formed and the amount of oxygen on the CNT surface *i.e.* initial CNT purity. These results can explain the wide discrepancy seen in transport behaviour of Ti-nanotube contacts found in the literature.

Although our results suggest ultra-high ($10^{-9}$ mbar) vacuums are necessary in order to obtain reproducible low resistance Ohmic Ti-nanotube contact, there may be ways in which lower vacuums could still be used. For example, the nanotubes should have their pre-existing oxygen content minimized via a pre-anneal, preferably once in the chamber. Oxygen contamination from oxide on the Ti source surface and residual gas in the chamber can be minimized by pre-deposition of sacrificial Ti away from the sample. Increasing the sputtering rate may also reduce exposure time to atmospheric oxygen. We emphasize that it is only oxygen at the Ti-nanotube interface which is important; later oxidation of the Ti surface should not affect contact properties.

The theoretical calculations predict mobility for Ti at temperatures not far above room temperature, so gentle surface heating during deposition (up to 100°C) may be sufficient to crystallize the Ti layer under vacuum (however Ti surface mobility is reliant on the absence of oxygen). A crystallized Ti layer should lead to significantly improved contact behaviour, and this is consistent with experimental studies of post-deposition heating which appear to promote TiC formation[22].

## Experimental Section

The samples were prepared using commercially available MWCNT powder synthesized by chemical vapour deposition[23]. Titanium was deposited onto the sample by e-beam evaporation from a Ti rod. The surface rod was cleaned of oxide by pre-sputtering onto a blanking plate. A quartz balance was used to calibrate the Ti evaporation amount. Samples 1-7 correspond respectively to 0.3, 1, 2, 5, 10, 20 and 100 Å nominal amounts of Ti deposition.

For the oxygen plasma pre-treated samples, the plasma treatment was carried out for MWCNTs on the carbon film support in order to avoid dispersion and post-treatment contamination. The plasma chamber is described elsewhere[24]. The treatment was performed inside the plasma glow discharge for 60 seconds with gas pressure of 0.1 Torr at 15 W[20].

TEM using a JEOL 3000F microscope operated at 300 kV was used to follow the metal deposition for increasing metal evaporated amounts. MWCNT produced by arc-discharge[25] were used for the TEM imaging. The MWCNT powder was dispersed in ethanol, and a drop of the solution was deposited on a lacey carbon film supported by a copper grid.

XPS measurements of the C *1s* and Ti *2p* core levels were performed at Hasylab at beamline BW2 using a photon energy of 3300 eV. The overall resolution of the system (source + analyzer) was 0.6 eV[26]. The Au *4f$_{7/2}$* peak at 84.00 eV, recorded on a gold sample reference was used for calibration of the binding energy scale. In order to check for any possible photon energy drift in the measurements, reference spectra were recorded before and after each core level data set recorded on all samples. To perform the XPS measurements, the MWCNT powder was pressed on a conductive adhesive tape suitable for ultra high vacuum. The thickness and the homogeneity of the CNT layer were checked to assure no interference of the tape on the measurements. From Figure 3 (inset) it can be seen that a porous CNT-film is formed. Additional XPS measurements reported in Figure 5 were performed using an SSI-100 spectrometer at a photon energy of 1486.6 eV.

Density Functional calculation[27,28] within the Local Density Approximation are carried out on a 8×8 supercell of graphene, i.e. a monolayer of 128 carbon atoms. Fully spin polarized single *k*-point calculations were geometrically optimized from multiple possible starting structures. Hartwigsen, Goedecker and Hutter (HGH) pseudopotentials[29] were used. Atom-centered Gaussian basis functions are used to construct the many-electron wave function with angular momenta up to $l$=2. Electronic level occupation was obtained using a Fermi occupation function with $kT$=0.04 eV. In the energetic analysis that follows, binding energies are defined as the difference in energy between the relaxed combined system and the isolated perfect graphite sheet and a single isolated metal atom unless stated otherwise. The diffusion barrier for Ti was determined by constraining the system symmetry and was confirmed as a saddle point by subsequent symmetry breaking and structural optimization which led to the ground state structure.

## Acknowledgements

*This work is financially supported by the Belgian Program on Interuniversity Attraction Pole (PAI 6/08), ARC-UMH , by the EU FP6 project STREP "nano2hybrids", Reference 003311 and by DESY and the European Commission under contract RII3-CT 2004-506008 (IASFS). JG is a research associate of NFSR (Belgium).*

**Keywords:** Carbon nanotubes · DFT · Photoemission spectroscopy · Titanium ·

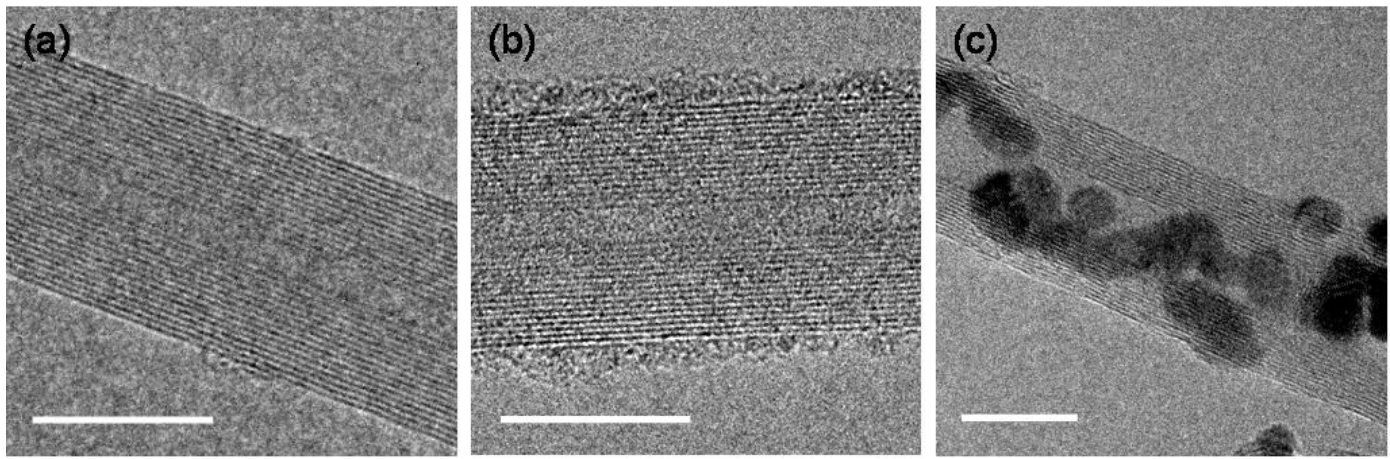

Figure 1. HRTEM micrographs of (a) pristine MWCNTs, (b) 10 Å of Ti on pristine MWCNTs and (c) 10 Å of Pd on pristine MWCNTs . Scale bars are 10 nm long

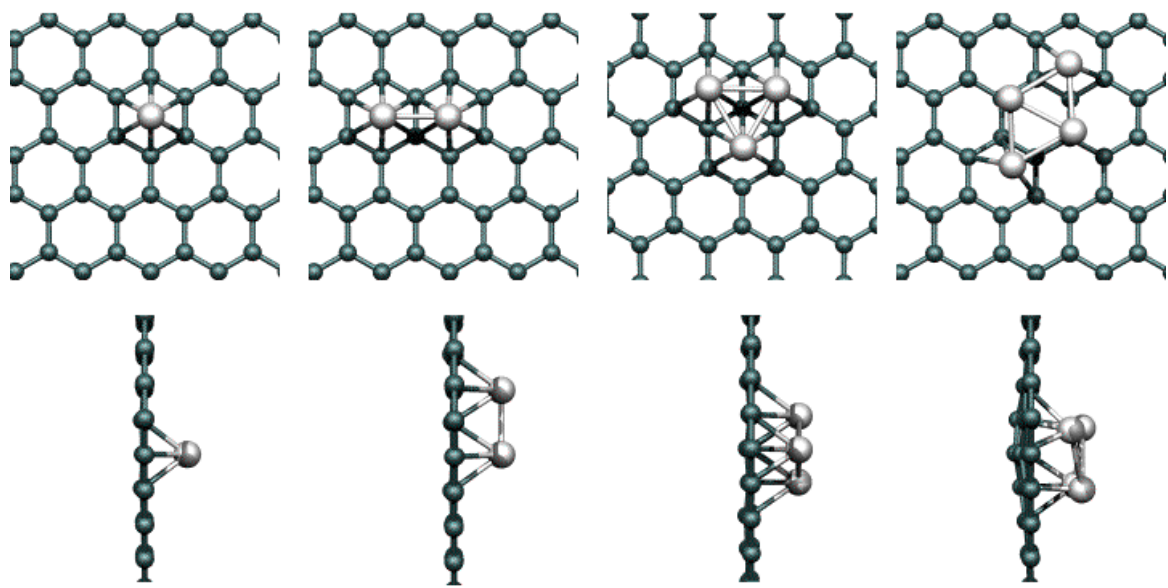

Figure 2. Top and side view of the DFT optimized structures of one to four Ti atoms on graphene

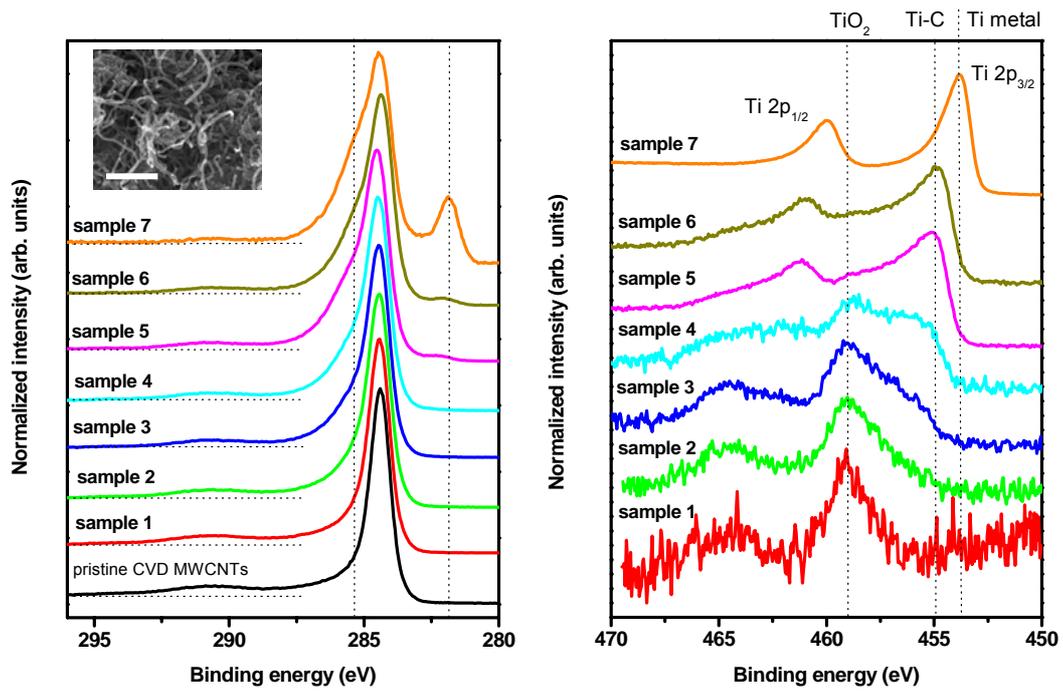

Figure 3. Evolution of C 1s (left) and Ti 2p (right) core levels for increasing coverage of titanium on pristine MWCNTs. Inset shows SEM image of the CNT film for Sample 7, i.e. after 10 nm Ti deposition (scale bar is 500 nm).

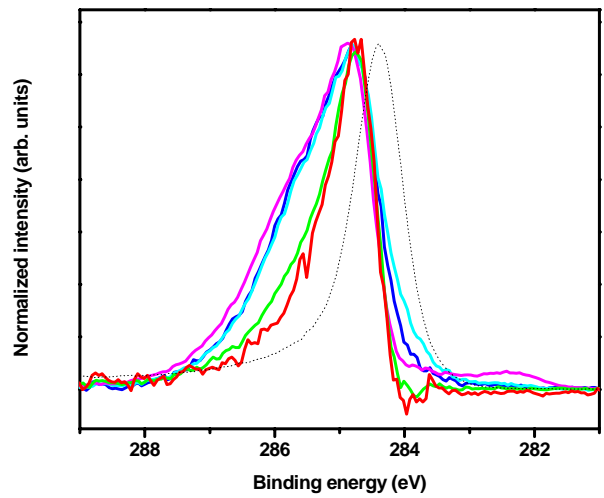

Figure 4. Subtracted XPS C 1s level of Ti decorated pristine nanotubes. The red, green, blue, cyan and magenta curves correspond respectively to sample 1, 2, 3, 4, 5. The C 1s level of pristine nanotube is shown for comparison (dotted curve).

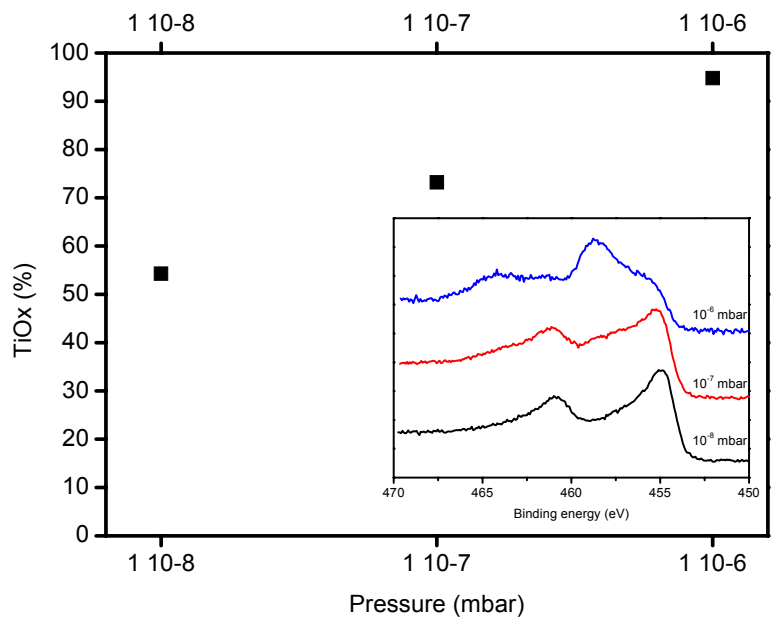

Figure 5. Titanium oxide content in a Ti layer evaporated on a gold substrate under different vacuum pressure. The insert shows the Ti 2p spectra of a 10 angstrom titanium layer deposited at 10-6, 10-7 and 10-8 mbar

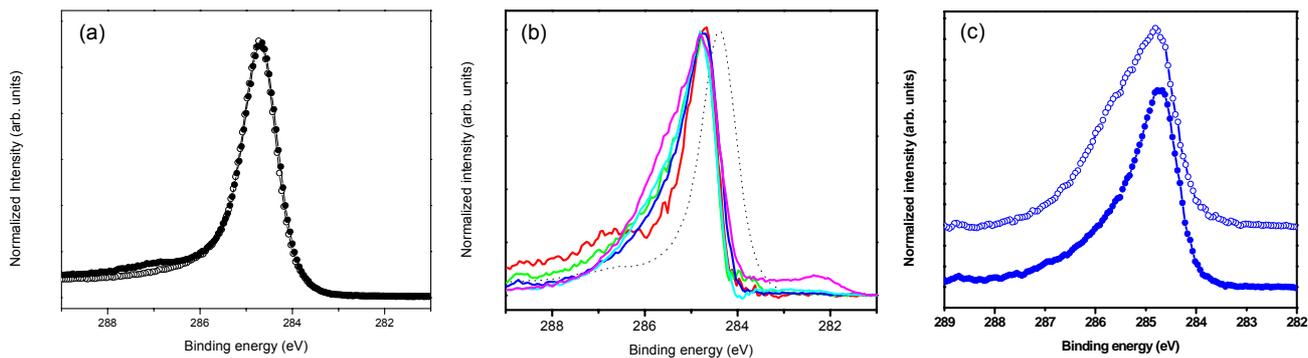

Figure 6. (a) C 1s spectra recorded from pristine and oxygen plasma treated MWCNTs (b) Subtracted XPS C 1s level of Ti decorated oxygen plasma treated nanotubes. The red, green, blue, cyan and magenta curves correspond respectively to sample 1, 2, 3, 4, 5. (c) Comparison between subtracted C 1s spectra of pristine (open circles) and oxygen plasma treated (solid circles) MWCNTs covered with 2 A of Ti.

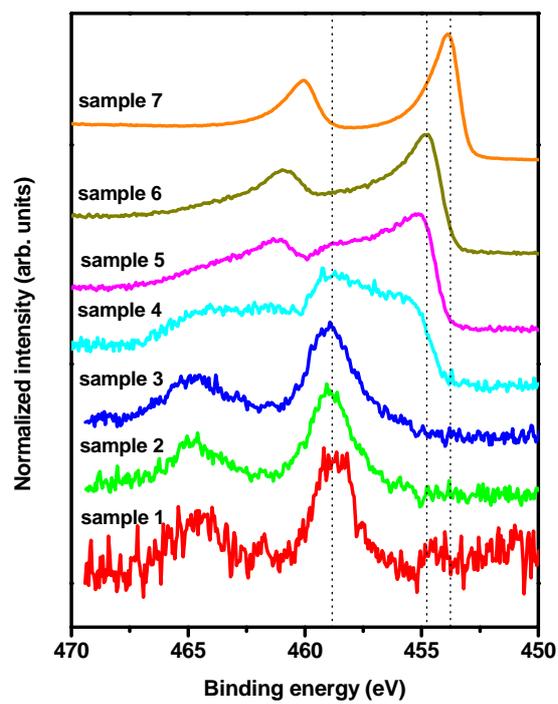

Figure 7. Evolution of Ti 2p core levels for increasing coverage of titanium on oxygen plasma treated MWCNTs

**Entry for the Table of Contents** (Please choose one layout)

Layout 1:

# ARTICLES

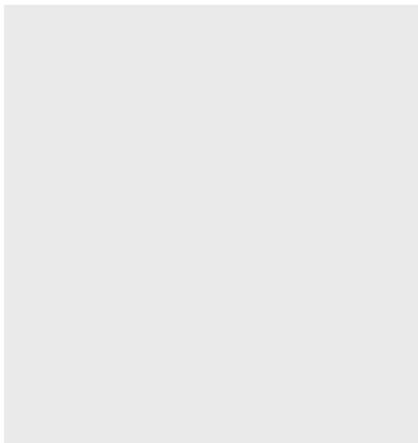

*Alexandre Felten, Irene Suarez-Martinez, Xiaoxing Ke, Gustaaf Van Tendeloo, Jacques Ghijsen, Jean-Jacques Pireaux, Wolgang Drube, Carla Bittencour]* and Christopher P. Ewels*

***Page No. – Page No.***

**The Interface between Titanium and Carbon Nanotubes Title**